\documentclass[11pt]{iopart}
\usepackage[english]{babel}

\expandafter\let\csname equation*\endcsname\relax 
\expandafter\let\csname endequation*\endcsname\relax 

\usepackage{amsthm}
\usepackage{amsmath}
\usepackage{latexsym}
\usepackage{amsfonts}
\usepackage{amssymb}
\usepackage{epstopdf}

\usepackage{graphicx}
\usepackage{booktabs}
\usepackage{verbatim}
\usepackage{dcolumn}
\usepackage{color}
\usepackage{xcolor}
\usepackage[normalem]{ulem}
\usepackage{soul}
\usepackage{braket}

\renewcommand{\thesection}{\arabic{section}}

\renewcommand{\thesubsection}{\thesection.\arabic{subsection}}

\begin{document}

\title{Retrieving past quantum features with deep hybrid classical-quantum reservoir computing}

\author{Johannes Nokkala}

\address{Department of Physics and Astronomy, University of Turku, FI-20014, Turun Yliopisto, Finland.}

\author{Gian Luca Giorgi}
\address{Instituto de F\'{i}sica Interdisciplinar y Sistemas Complejos (IFISC, UIB-CSIC), Campus Universitat de les Illes Balears E-07122, Palma de Mallorca, Spain.}
\author{Roberta Zambrini}
\address{Instituto de F\'{i}sica Interdisciplinar y Sistemas Complejos (IFISC, UIB-CSIC), Campus Universitat de les Illes Balears E-07122, Palma de Mallorca, Spain.}
\ead{roberta@ifisc.uib-csic.es}
\begin{abstract}
Machine learning techniques have achieved impressive results in recent years and the possibility of harnessing the power of quantum physics opens new promising avenues to speed up classical learning methods. Rather than viewing classical and quantum approaches as exclusive alternatives, their integration into hybrid designs has gathered increasing interest, as seen in variational quantum algorithms, quantum circuit learning, and kernel methods. Here we introduce deep hybrid classical-quantum reservoir computing for temporal processing of quantum states where information about, for instance, the entanglement or the purity of past input states can be extracted via a single-step measurement.  We find that the hybrid setup cascading two reservoirs not only inherits the strengths of both of its constituents but is even more than just the sum of its parts, outperforming comparable non-hybrid alternatives. The quantum layer is within reach of state-of-the-art multimode quantum optical platforms while the classical layer can be implemented in silico.
\end{abstract}

\maketitle

\section{Introduction}

Classical machine learning techniques have shown their outstanding impact in handling vast datasets, while quantum machine learning offers groundbreaking potential by harnessing the power of quantum mechanics to revolutionize problem-solving with enhanced computational capabilities. This has spurred an interest in hybrid classical and quantum machine learning techniques for quantum machine learning  \cite{hinton2006fast,Goodfellow2016,farhi2014quantum,kandala2017hardware,mitarai2018quantum,zhu2019training,verdon2019learning,mujal2021opportunities,cerezo2021variational,bharti2022noisy,ghukasyan2023quantum}. This innovative approach harnesses the specific capabilities of both classical and quantum systems, leveraging classical algorithms and quantum computational power to address complex problems that surpass the limits of classical methods alone.  Popular implementations are variational quantum algorithms \cite{cerezo2021variational} used in optimization and learning problems that combine parameterized quantum circuits implementable in noisy intermediate scale quantum (NISQ) devices \cite{bharti2022noisy} with classical optimization techniques. Examples are quantum variational eigensolvers \cite{peruzzo2014variational,kandala2017hardware}, quantum approximate optimization algorithms \cite{farhi2014quantum}, generative models \cite{zhu2019training}, and classifiers \cite{Chen_2021}. The possibility of combining classical neural networks to optimize control parameters of parametrized quantum circuits has also been explored in the context of meta-learning \cite{verdon2019learning} and quantum circuit learning \cite{mitarai2018quantum}. Recently, multiple Kernel methods have also been proposed \cite{ghukasyan2023quantum} in (hybrid) pairwise combinations of several classical-classical, quantum-quantum, and quantum-classical kernels with support-vector machines to improve their expressivity. An architecture that naturally falls within the definition of hybrid classical-quantum algorithms is represented by quantum reservoir computing (QRC) \cite{PhysRevApplied.8.024030,martinez2021dynamical,tran2021learning}, together with its static version known as quantum extreme learning machine \cite{ghosh2019quantum,PhysRevLett.123.260404,suprano2023experimental,Krisnanda,Innocenti2023} (see Ref. \cite{mujal2021opportunities} for an inclusive review). QRC combines the richness and expressivity of the quantum dynamics with the easy and fast trainability typical of classical reservoir computing, an approach for time-series processing inspired by recurrent neural networks and where it suffices to optimize the output layer leaving the hidden layer(s) untouched \cite{jaeger2001echo,maass2004computational,nakajima2021reservoir}.

Meanwhile, deep neural networks have shown their outstanding capabilities \cite{hinton2006fast,Goodfellow2016}. The idea behind such deep learning is that complex problems can be better solved by taking advantage of the ability to process data hierarchically, addressing the different characteristics of the inputs in each of the different layers within the model. Deep learning has been applied very successfully to, for instance, computer vision or natural language processing \cite{voulodimos2018deep,young2018recent} as well as detection of entanglement from incomplete measurement data \cite{koutny2023deep}. One of the main issues with deep learning is the large number of parameters that need to be optimized in the training phase, which makes it extremely costly. This problem becomes much less significant when deep architectures meet reservoir computing algorithms where only the output layer needs to be trained. Examples of deep reservoir computing can be found in Refs.~\cite{triefenbach2010phoneme,gallicchio2017deep,freiberger2019improving,nakajima2022physical,lin2022deep,lupo2023deep} where it was shown that the sequential structure of the reservoir hidden layers enriches the dynamics and, as a consequence, improves the performance of the system.

In this work,  we combine the pursuit of hybrid approaches and deep learning by introducing deep classical-quantum reservoir computing. In our proposal, the deep architecture is hybrid itself, as the hidden layers consist of a cascade of a QRC and a classical echo state network (ESN). QRC allows the coherent processing of quantum inputs for temporal tasks \cite{nokkala2023online}, has the potential for a rapidly increasing information processing capacity due to its rich dynamics \cite{mujal2021opportunities}, and also inherits many of the benefits of its classical counterpart \cite{nakajima2022physical}: engineering simplicity, multitasking, and fast training. Indeed, the internal connections of the reservoir can be random and a fixed reservoir can solve many tasks by just using a different output layer trained by simple linear regression. 

Reservoir computing has been successfully implemented in different platforms for time series processing \cite{tanaka2019recent}. When moving to QRC implementations, the output extraction presents new challenges due to the invasive nature of quantum measurements \cite{mujal2021opportunities}. Interestingly, decoherence does not always represent a limiting factor \cite{chen2020temporal,sannia2022dissipation,PhysRevResearch.5.023057,domingo2023taking}. Sequential processing can be achieved with different strategies based on indirect  \cite{mujal2023time,garcia2023scalable,garcia2023squeezing} and quantum non-demolition measurements \cite{yasuda2023quantum}.
In particular,  Refs. \cite{garcia2023scalable,garcia2023squeezing} combine input injection, output extraction, and a physical ensemble in a scalable multimode quantum optics platform.  Real-time QRC is performed continuously on the injected quantum states, with sequential processing not relying on external memories.  Furthermore, 
this implementation is successful with experimentally convenient resources, e.g.  Gaussian states. 

Building on the advances in both classical and quantum approaches, our goal is to show the advantage of their integration into hybrid designs. Indeed, while quantum settings allow for a natural quantum input embedding, the output observables are limited to linear temporal processing \cite{innocenti}. Additionally, for any finite ensemble, they suffer from rapidly decaying memory \cite{mujal2023time,garcia2023scalable} and expressivity \cite{tureciPRX23}, being the signal-to-noise ratio of the observables a limiting factor. On the other hand, considering only a classical reservoir computer injected with a (measured) quantum input would require exponential resources or not reveal full information of the data without suitable quantum preprocessing (e.g. when the measurements are not tomographically complete).

We show that our hybrid setup combining the real-time QRC \cite{garcia2023scalable} with a powerful and versatile classical reservoir computer, namely an echo state network \cite{jaeger2001echo}, has the same strengths as both of its constituents. As opposed to the case of standard QRC \cite{Innocenti2023}, nonlinear functionals of the temporal quantum data become possible.  Additionally, new strengths appear as the memory can overcome the decay of the quantum layer alone, and significant advantages over purely quantum and purely classical alternatives may be observed in both linear and nonlinear tasks. In the proposed hybrid architecture, the quantum preprocessing allows the reconstruction of the full input covariance matrix from incomplete (homodyne) measurement. The one-time output reveals many past input states or their quantum features (for instance entanglement). With the QRC, high-fidelity reconstruction would require large ensembles with a size that needs to increase exponentially with the delay to sustain the signal-to-noise ratio (SNR) of all the state components. The ESN can assist in an accurate reconstruction by post-processing the measurement data. Indeed, this can reduce the required ensemble size by over an order of magnitude. We show how the nonlinear memory allows the detection of, for example, entanglement and purity, provided both the real-time QRC and ESN are of moderate size. 

\section{Methods}

\begin{figure}
\centering
    \includegraphics[trim=0cm 9cm 15.7cm 0cm,clip=true,width=0.45\textwidth]{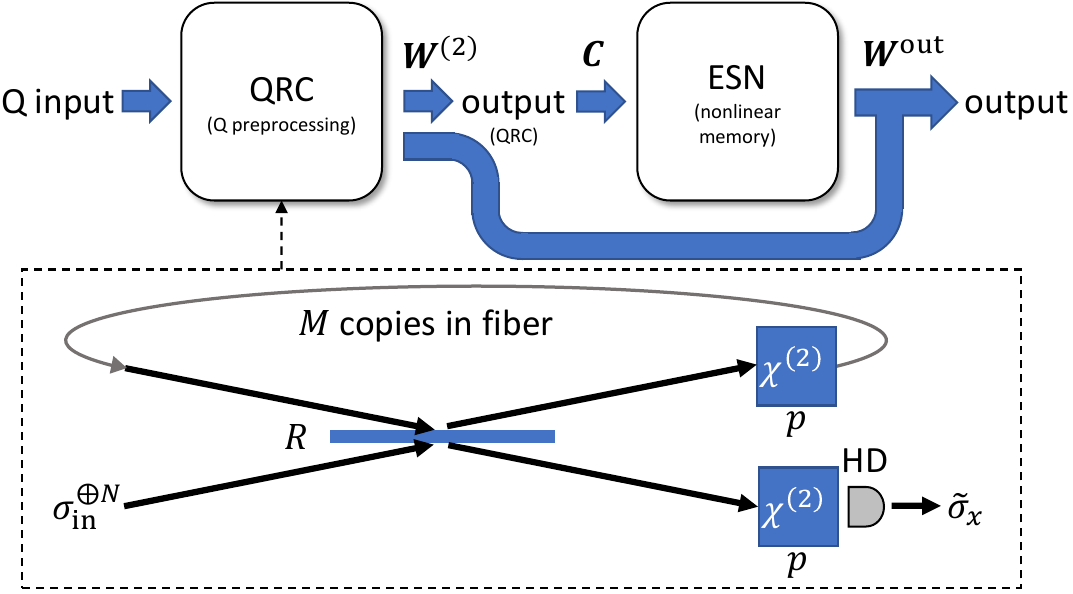}
    \caption{
    The hybrid architecture. The input, a time series of quantum states, is processed by the QRC to produce a time series of classical information as a linear combination of its estimated observables according to the trained matrix $\mathbf{W}^{(2)}$. This is injected to the ESN through a random coercion matrix $\mathbf{C}$. The final output is a trained linear combination of both the estimated QRC observables and ESN variables according to matrix $\mathbf{W^{\mathrm{out}}}$. Specifically, at each time step a product state of $N$ copies of the input interacts with the quantum reservoir---a physical ensemble of size $M$---via a beam splitter of reflectivity $R$. The two nonlinear $\chi^{(2)}$ crystals with sparsity $p$ couple the optical modes both in the optical fiber loop and before the homodyne measurements. The used observables are the elements of the maximum likelihood estimate of the covariance matrix of the $x$-quadratures, $\tilde{\sigma}_x$.
    }
    \label{fig:scheme}
\end{figure}

Reservoir computing is an established supervised approach for temporal tasks, where each output at a given time can be trained to approximate any functional of the recent input history \cite{nakajima2021reservoir,konkoli2017reservoir,tanaka2019recent}. Quantum reservoirs, beyond potential performance boosts, offer the unique advantage of coherent processing of quantum states as inputs  \cite{mujal2021opportunities,tran2021learning,nokkala2023online,de2023quantum}, as will be done here. While quantum states can be processed by classical computers, this is generally exponentially costly in terms of resources. In this sense, using a quantum reservoir offers an intrinsic advantage. 
Our proposal, shown in Fig.~\ref{fig:scheme}, is based on combining a quantum and a classical reservoir computer in a deep or cascaded configuration. The goal is to benefit both from the quantum preprocessing and the nonlinear classical post-processing to solve {quantum} tasks efficiently, with high accuracy and with experimentally convenient resources. The classical RC consists of an ESN (described below). The output of the hybrid classical-quantum reservoir computer is trained to the desired tasks, optimizing the matrices $\mathbf{W}^{(2)}$ and $\mathbf{W}^{\mathrm{out}}$, which form the QRC output and the ESN output, respectively (see Fig. \ref{fig:scheme}).

The quantum layer is accounted for by the real-time QRC introduced in Ref.~\cite{garcia2023scalable}, based on a multimode quantum optics platform shown in the bottom half of Fig.~\ref{fig:scheme}. The reservoir consists of  $N$ optical modes in a feedback loop and coupled {to each other} by a nonlinear $\chi^{(2)}$ crystal. A physical ensemble of $M$ identical copies of reservoirs can be implemented as a sequence circulating inside an ideal lossless optical fiber. The input modes are in zero mean Gaussian states and interact with the reservoir modes through a beam splitter of reflectivity $R$ such that at $R=1$ they are reflected directly to the readout---a second $\chi^{(2)}$ crystal followed by homodyne measurements of the $x$-quadratures of all modes---and at $R=0$ will be transmitted, circulate once and then transmitted again to the readout. Nontrivial memory effects appear at intermediate values of $R$ where each reflected set of modes depends on (recent) input history, with memory provided by the feedback loop. The nonlinear crystals distribute the information across the available degrees of freedom, leading to linearly independent observables that play the role of the computational nodes.

The Hamiltonian of a $\chi^{(2)}$ crystal in the QRC is of the form
\begin{equation}
\begin{split} 
        H_{\chi^{(2)}} &= \sum_{j=1}^{N} \omega_{j} \left(a^{\dagger}_{j} a_{j}+\frac{1}{2}\right) \\
        &+ \sum_{l>j} \left(g_{jl}a^{\dagger}_{j} a_{l} + \mathrm{i} h_{jl} a^{\dagger}_{j} a^{\dagger}_{l} + \mathrm{h.c.} \right) ,
\end{split}
\end{equation}
where $\mathrm{i}=\sqrt{-1}$, $a_j$ and $a_j^\dagger$ are the annihilation and creation operators, respectively, indexed by mode and satisfying the commutator relation $[a_j,a_l^\dagger]=\delta_{jl}$. The reservoir is a complex network \cite{nokkalareview,renaultPRXquantum} with (real) couplings $g_{jl}$ and $h_{jl}$. Following Ref.~\cite{garcia2023scalable} they are random numbers uniformly distributed as $g_{jl}\in[0.1,0.3]$ and $h_{jl}\in[0.2,0.4]$, scaled with the modes frequencies ($\omega_j=1$); these parameters are consistent with squeezing levels of state-of-the-art experiments. The time $\Delta t$ inside each crystal is set to one. Here we additionally introduce a sparsity parameter $p$ which controls whether a coupling is on or off in the network, that is to say once all weights have been randomized each is individually kept with probability $p$ and otherwise set to $0$ such that still $g_{jl}=g_{lj}$ and $h_{jl}=h_{lj}$. For $p=1$ the network is generally fully connected, whereas setting $p=0$ removes all couplings and effectively the crystals. As shown in Appendix~A, the fully connected configuration is not always the best option as one could expect, and intermediate values can optimize performance.

Because all states will be zero mean Gaussian states and the Hamiltonians are quadratic in the creation and annihilation operators, the relevant state remains a multimode squeezed vacuum state at all times and as such is fully described by its covariance matrix
$ (\sigma(\mathbf{x}))_{jl}=\langle\mathbf{x}_j\mathbf{x}_l+\mathbf{x}_l\mathbf{x}_j\rangle/2-\langle\mathbf{x}_j\rangle\langle\mathbf{x}_l\rangle$
where $\mathbf{x}=\{x_1,p_1,x_2,p_2,\ldots\}^\top$ is the vector of the quadrature operators of the field indexed by mode. In general, $x=a+a^\dagger$ and $p=(a-a^\dagger)/\mathrm{i}$, making the covariance matrix of the vacuum state $\Ket{0}$ normalized such that $\sigma(x,p)=\big(\begin{smallmatrix}
  1 & 0\\
  0 & 1
\end{smallmatrix}\big)$.
For each time step $k$, the input is a product state \cite{garcia2023scalable,garcia2023squeezing}, with the covariance matrix $\sigma(\mathbf{x}_{k}^{\mathrm{in}})^{\oplus N}$ where $\oplus$ is the direct sum (for the sake of reducing clutter we drop the exponent from now on). With each injection, the input quantum state is set to a new value and coupled to the cavity modes through the beam splitter (bottom of Fig.~1). Therefore, at each time step $k+1$ there are not initial correlations between the reservoir and input modes, $\sigma(\mathbf{x}_{k}^R\oplus\mathbf{x}_{k+1}^{\mathrm{in}})=\sigma(\mathbf{x}_k^R)\oplus\sigma(\mathbf{x}_{k+1}^{\mathrm{in}})$. Here $\mathbf{x}_{k}^R$ accounts for the reservoir modes at the end of the previous time step, $\mathbf{x}_{k+1}^{\mathrm{in}}$ are the new input modes. The covariance matrix of the reservoir and output updates as
\begin{equation}
\sigma(\mathbf{x}_{k+1}^R\oplus\mathbf{x}_{k+1}^{\mathrm{out}})=\mathbf{S}(\Delta t)\sigma(\mathbf{x}_{k}^R\oplus\mathbf{x}_{k+1}^{\mathrm{in}}))\left(\mathbf{S}(\Delta t)\right)^\top
\label{eq:QRC}
\end{equation}
where the $4N\times4N$ symplectic matrix $\mathbf{S}(\Delta t)$ accounts for the action of the beam splitter and the two nonlinear $\chi^{(2)}$ crystals (see Eq.~$(B5)$). {Then, due to the complex dynamics of the reservoir, $\sigma(\mathbf{x}_{k+1}^R\oplus\mathbf{x}_{k+1}^{\mathrm{out}})$ in general displays correlations; we remind that the argument $\mathbf{x}_{k+1}^R\oplus\mathbf{x}_{k+1}^{\mathrm{out}}$ is just a vector of operators.} The QRC outputs are real numbers obtained from the homodyne measurements of the $x$-quadratures of the output modes, and beyond ideal conditions, we consider finite ensembles (of size $M$). Specifically, the output $o_{k+1}^{\mathrm{QRC}}$ is a linear combination of the elements of $\tilde{\sigma}_x(\mathbf{x}_{k+1}^{\mathrm{out}})$ and a bias term. Here $\sigma_x$ is the $N\times N$ covariance matrix of only the $x$-quadratures and $\tilde{\sigma}_x$ its maximum likelihood estimate calculated from the homodyne measurement data over the ensemble of size $M$; because $\tilde{\sigma}_x$ is a symmetric matrix it provides $N(N+1)/2$ computational nodes. {In the case at hand the measurement back-action averages out as explained in the Appendices of Ref.~\cite{garcia2023scalable}.} In QRC, the output layer is usually easily trained through a Ridge regression actually implemented on a classical computer \cite{mujal2021opportunities}, so that a somewhat hybrid nature of the architecture is already present in the original setting  \cite{pfeffer2022hybrid}. Here instead we consider a deep hybrid design where both the quantum and the classical components are reservoir computers, in order to boost the overall performance and enquire about their respective capabilities.

The classical layer is an ESN, a powerful and versatile reservoir computer \cite{jaeger2001echo}. In practice, it would be implemented in silico on a (classical) computer. The state of the ESN at some timestep $k$ is a vector of $N_{\mathrm{ESN}}$ real numbers $\mathbf{x}_{k}^{\mathrm{ESN}}$ where $N_{\mathrm{ESN}}$ is the number of neurons in the network. With each new input, the state updates as
\begin{equation}    \mathbf{x}_{k+1}^{\mathrm{ESN}}=f\left(\rho\mathbf{W}\mathbf{x}_{k}^{\mathrm{ESN}}+\iota\mathbf{C}\mathbf{s}_{k+1}\right),
\label{eq:ESN}
\end{equation}
where $f$ is the activation function, $\rho,\iota\in\mathbb{R}$ are the feedback gain and input gain parameters controlling the importance of the previous state and latest input, respectively.
The classical ESN vector $\mathbf{x}_{k+1}^{\mathrm{ESN}} $ and the (new) injected classical inputs $\mathbf{s}_{k+1}$ (vector of dimension $d$ coming form  the covariance outputs of the QRC) are linearly combined in this classical reservoir through two random real matrices $\mathbf{W}$ and  $\mathbf{C}$. The weight $N_{\mathrm{ESN}}\times N_{\mathrm{ESN}}$ matrix $\mathbf{W}$ accounts for the internal connections between the ESN neurons; the elements from a uniform random distribution in the interval $[-1,1]$, are scaled setting a unit spectral radius of the matrix. The $N_{\mathrm{ESN}}\times d$ coercion matrix $\mathbf{C}$ is generated in the same way and is responsible for converting the input to match the ESN size. The activation function is $f(x)=\ln(1+\exp(x))$ and the function acts elementwise. The hybrid architecture output $o_{k+1}$ is a linear combination of all computational nodes, namely the $N(N+1)/2$ independent elements of $\tilde{\sigma}_x$, the $N_{\mathrm{ESN}}$ variables in $\mathbf{x}_{k+1}$, and a bias term.

As common in reservoir computing, the sufficiently long input time series is divided into three parts, namely preparation (or wash-out), training, and test. During preparation, the transient dependency on the initial state is removed (echo state property). Training aims to optimizing $\mathbf{W}^{(2)}$ and then $\mathbf{W}^{\mathrm{out}}$ (see top of Fig.~1), in a two-step process. The elements of $\tilde{\sigma}_x$ are recorded during the training phase and linear regression is used to minimize the squared error between a given target output and the actual output, fixing $\mathbf{W}^{(2)}$. This output is used as input by the ESN and the process is repeated to optimize the weights in $\mathbf{W^{\mathrm{out}}}$. Finally, the test phase is used to assess how well the trained reservoir computer generalizes to cases not used in training; all the shown results in the following represent tasks performance for the unseen data, i.e. in the test phase (see also Appendix~C). 
We note that in this two-step process, the target for training $\mathbf{W}^{(2)}$ needs not be the same as the final target. We will describe all targets as the considered tasks are introduced.

The performance of our hybrid architecture will be tested in four different tasks, displaying the ability of the system to store and process information of random sequences of quantum states: short-term memory, as well as trace, determinant, and entanglement detection with delay. The first two require linear memory to be efficiently solved, while the last two are nonlinear.

\begin{table}[h]
\caption{Relevant hyperparameters in QRC and ESN, and default values used in benchmark tasks. Feedback gain $\rho$ and input gain $\iota$ have different values for linear {(short-term memory, trace)} and nonlinear {(determinant, entanglement detection)} tasks.
}
\centering
\begin{tabular}{ p{1.1cm}p{4.95cm}p{2cm} }
\hline
 \hline
 Symbol& \multicolumn{1}{c}{Meaning} &Default value\\
 \hline
$N$   & Number of optical modes    &9\\
$R$&   Reflectivity  & 0.4\\
 $p$ &Sparsity parameter & $7/9$\\
 $M$    &Ensemble size & $10^5$\\
$N_{\mathrm{ESN}}$&   Number of neurons  & 45\\
 $\rho$& Feedback gain  & $0.7$ (lin.) \; ~~ $0.1$ (nonlin.)   \\
 $\iota$& Input gain  & $10^{-4/3}$ (lin) \;  $0.01$ (nonlin.)\\  
 $\tau$& Delay  & --\\
 $\tau'$& Delay for real-time QRC & $\lceil\tau/2\rceil$\\
 --& Length of preparation phase  & 500\\
 --& Length of training phase  & 3000\\
 --& Length of test phase  & 1000\\ 
 \hline
\hline
\end{tabular}
\label{tab:parameters}
\end{table}

For convenience, the selected parameters for the presented results, when not differently indicated, are reported in Tab.~\ref{tab:parameters}. How their values have been chosen is explained in Appedix~A. In particular, the default values of $N$ and $N_{\mathrm{ESN}}$ are chosen to ensure an equal distribution of computational nodes between the quantum and classical layers. 

\section{Results}

\subsection{Memory of quantum states}

Let us begin our analysis by assessing the reservoir computer's ability to recall past (quantum) inputs through the short-term memory task, a standard benchmark in classical RC here generalized to quantum state sequences. The target for the deep hybrid setting is to output at time $t$ the full information of a quantum state injected at time ${t-\tau}$. Let $\sigma_t$ be the short-hand for the input covariance matrix at time step $t$; then the final target is simply $\sigma_{t-\tau}$ where $\tau\geq 0$ is a delay. In the proposed hybrid architecture, the QRC is trained to output the unique elements of the input covariance matrix of Gaussian states with a delay $\tau'\leq\tau$. The figure of merit is fidelity $F$ (see Eq.~$(B7)$) between the input state (injected in the past) and the approximation achieved by the output after a delay time delay $\tau$. As the output usually has some errors it is not guaranteed that $F\in[0,1]$ as it should; such unphysical cases are very rare especially when the performance is good, and are therefore simply excluded.  

We start considering, as inputs, random zero mean squeezed thermal states with a covariance matrix distributed as
\begin{equation}                \sigma(n_{\mathrm{th}},r,\varphi)\sim\sigma(\mathcal{U}(0,5),\mathcal{U}(0,0.75),\mathcal{U}(0,2\pi))
\label{eq:singlemodeinput}
\end{equation}
where $\mathcal{U}(\mathrm{min},\mathrm{max})$ is a continuous uniform distribution in the closed interval $[\mathrm{min},\mathrm{max}]$ and with a slight abuse of notation we indicate how the thermal excitations $n_{\mathrm{th}}$, squeezing parameter $r$ and phase of squeezing $\varphi$ are distributed.

\begin{figure}[t]
\centering
    \includegraphics[trim=0cm 0cm 0cm 0cm,clip=true,width=0.45\textwidth]{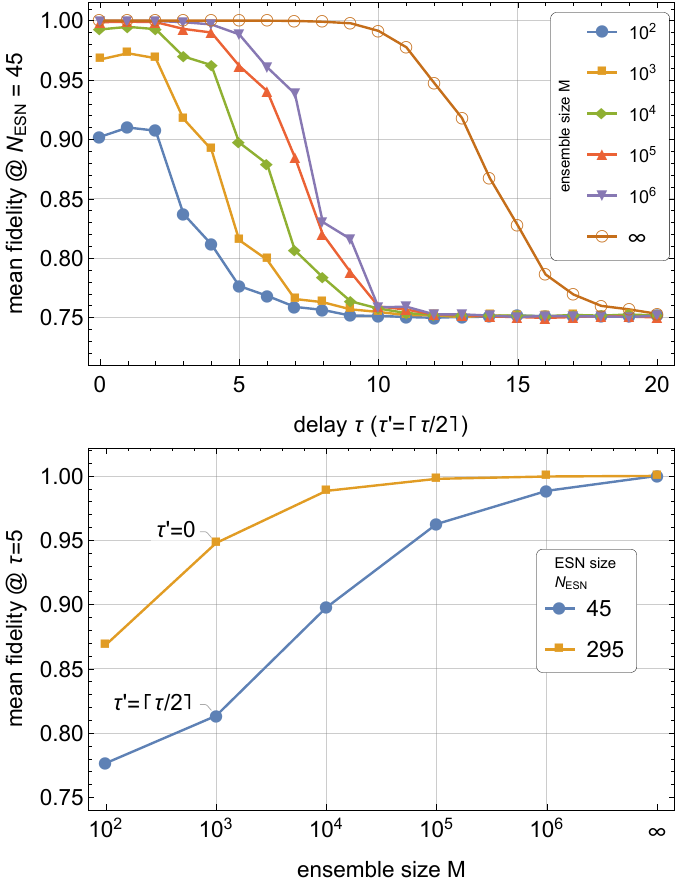}
    \caption{Effect of the ensemble size $M$ on mean performance over $100$ random realizations in input state estimation task and enhancement of the performance with a larger ESN. On the top panel, both the QRC and the ESN contribute with 45 trained weights and the delay $\tau$ is split evenly such that the real-time QRC is trained for delay $\tau'=\lceil\tau/2\rceil$. For smaller ensembles, the performance drops quickly as $\tau$ increases. On the bottom panel, the delay is fixed to $\tau=5$ and the default settings are compared to the hybrid setup with a larger ESN with $N_{\mathrm{ESN}}=295$ with increased feedback gain $\rho=1.8$ and $\tau'=0$. All other parameters are as in Tab.~\ref{tab:parameters}.}
    \label{fig:bigMbenchmark}
\end{figure}

One of the features of the real-time QRC, beyond ideal assumptions, is that for any finite $M$ the observables have limited SNR \cite{mujal2023time,garcia2023scalable}. The exact ideal covariance and averages of observables can be determined only at the limit $M=\infty$. The achievable performance for finite ensembles decreases exponentially with delay \cite{garcia2023scalable}, limiting the short-term memory and affecting also the hybrid reservoir computer. This is investigated in the upper panel of Fig.~\ref{fig:bigMbenchmark}. As expected, there is quite a large memory gap between the ideal case $M=\infty$ and a large ensemble of $M=10^6$ especially at longer delays. Remarkably, in the proposed hybrid RC, the classical and quantum reservoir delays can be adjusted independently to sum up the total delay $\tau$. This opens the possibility of improving the performance in memory tasks. Indeed in the lower panel of Fig.\ref{fig:bigMbenchmark}, the delay is fixed to $\tau=5$ and the default parameters are compared to the case where short-term memory is boosted by increasing the size and feedback gain of the ESN while setting the QRC delay $\tau'=0$. This greatly reduces the physical overhead required to reach a given fidelity: for instance, with bigger ESN the fidelity is $F\approx 0.95$ at $M=1000$ but in the default case this requires an ensemble of size $M=10^5$. Importantly, measurements of only the $x$-quadratures would not normally suffice for high fidelity for any delay as they would not determine the entire single-mode covariance matrix. However, this limitation is lifted by the QRC. Indeed, the  {$\chi^{(2)}$ crystals} enable interactions in the reservoir that effectively distribute the information across different optical modes.  A more complete state reconstruction is then possible through the observables of the optical network.

\subsection{Linear and nonlinear memory for energy, purity and entanglement retrieval\label{sec:tasks}}

\begin{figure}
\centering
    \includegraphics[trim=0cm 0cm 0cm 0cm,clip=true,width=0.95\textwidth]{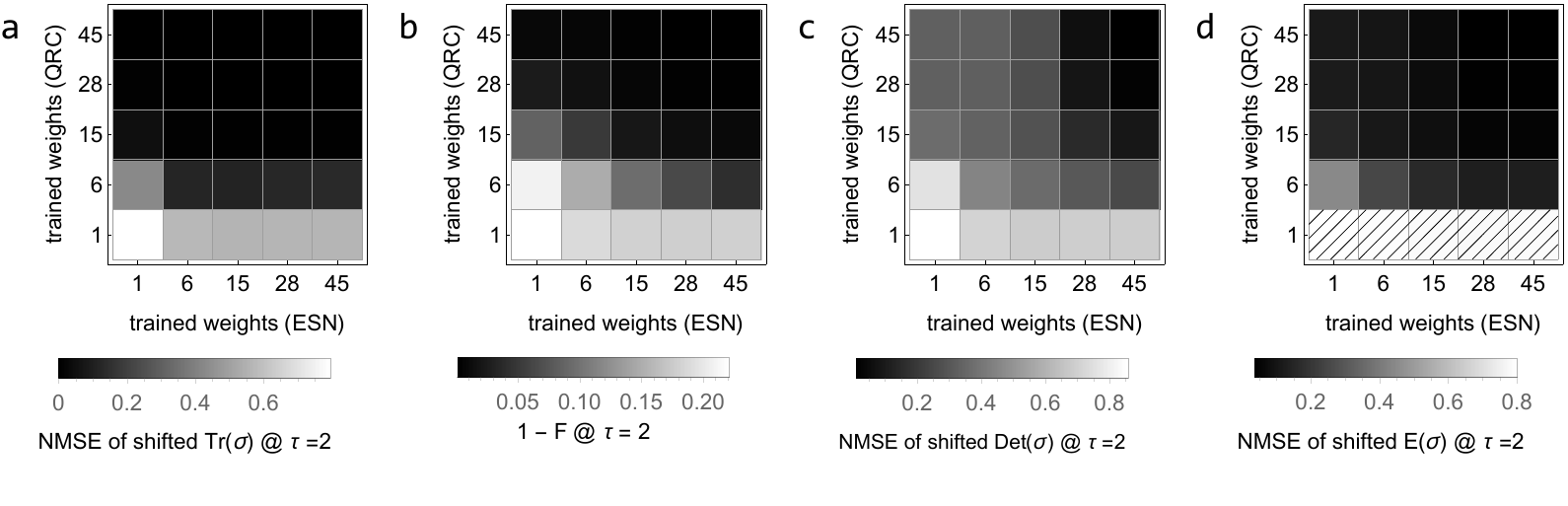}
    \caption{Effect of the reservoir size on mean performance over $100$ random realizations in linear (shifted trace of panel $(a)$ and state estimation of panel $(b)$) and nonlinear (determinant of panel $(c)$ and entanglement estimation of panel $(d)$) tasks. Details of the tasks are given in Sec.~\ref{sec:tasks}---in particular, the last one is not well defined for the smallest real-time QRC size (hatched line). The delay is fixed to $\tau=2$ and split evenly between QRC and ESN in the deep architecture. Performance is quantified by NMSE except in state estimation task where it is quantified by infidelity. 
    All parameters are as in Tab.~\ref{tab:parameters}.}
    \label{fig:lattices}
\end{figure}

As the QRC dynamics is governed by a linear input-output map for the state \cite{Innocenti2023}, Equation~\eqref{eq:QRC} is linear in the input covariance matrix. The hybrid setup, instead, has nonlinear memory thanks to the ESN. Here we investigate both the linear and nonlinear memory and check how performance depends on both $N$ and $N_{\mathrm{ESN}}$. To this end, we introduce another simple linear task and two nonlinear tasks to provide a broad assessment. 
Specifically, as a linear task in the quantum state, we consider the memory trace task where the target is {the trace of the covariance matrix} with delay, $\mathrm{Tr}(\sigma_{t-\tau})$, proportional to the input energy -- and not to be confused with the trace of the density operator. The target for the real-time QRC in the trace task is $\mathrm{Tr}(\sigma_{t-\tau'})$.

More complex tasks are the retrieval of entanglement and  purity of past quantum inputs whose information is stored in the reservoir, based on a (delayed) single-time measurement. Both entanglement and purity are nonlinear tasks, obtained respectively from the lower symplectic eigenvalue and from the determinant of the covariance matrix. Their retrieval, in presence of delay,  
requires nonlinear memory. The respective targets for the QRC are set as in the short-term memory task and the final targets of the deep reservoir are $\mathrm{Det}(\sigma_{t-\tau})$ and the logarithmic negativity $\mathrm{E}(\sigma_{t-\tau})$, whose definition is reminded in Appendix~B. All of the new tasks reveal partial information about the input: as explained in Appendix~B, the trace of a covariance matrix is directly proportional to the energy of the system (since the mean is zero), the determinant completely determines the purity, and the logarithmic negativity is an entanglement measure \cite{vidal2002computable}. Unlike for others, for the entanglement detection task, the input states are two-mode squeezed thermal states distributed as 
\begin{equation}                \sigma(n_{\mathrm{th}},s)\sim\sigma(\mathcal{U}(0,1.25),\mathcal{U}(0,0.75))
\label{eq:2modeinput}
\end{equation}
where $n_{\mathrm{th}}$ is the amount of initial thermal excitations in either mode and $s$ is the two-mode squeezing parameter. 

We address the performances in all these tasks, assessing the improvement provided by increasing the sizes of the quantum and classical reservoir layers and results are summarized in Fig. \ref{fig:lattices}.The figure of merit for delayed energy, purity, and entanglement is the NMSE (see Eq.~$(C2)$), which vanishes for perfect correlation between the target and actual output and becomes $1$ at the limit of a long training phase when there is no correlation \footnote{Notice that the targets are shifted to have zero mean. Otherwise, the bias term alone could achieve $\mathrm{NMSE}<1$ by simply giving the correct mean. While this shifting can only decrease the observed NMSE, it also ensures that it reflects strictly the ability of the reservoir to perform the task.}. For the state estimation task infidelity $1-F$ is shown. Results in Fig.~\ref{fig:lattices} display all these tasks performance when  $N$ varies from $1$ to $9$ in steps of $2$ (and the number of trained weights label the axis). Notice that in the delayed entanglement detection, if $N$ is even, $N/2$ copies of the input state \ref{eq:2modeinput} are introduced at each time step, and if it is odd the extra mode will be in the vacuum state; consequently the task is not well defined for $N=1$ (as indicated by the hatching in the panel of the last panel of Fig. \ref{fig:lattices}). For ESN the number of trained weights coincides with network size $N_{\mathrm{ESN}}$. In all cases the quantum layer is necessary and cannot be compensated by the ESN. While the performance of the classical reservoir is always improved by the quantum layer, the latter can pose a bottleneck, as is particularly evident in the trace task. However already with $N=3$ (6 trained wigths in the QRC) a decent performance can generally be reached if the ESN size is sufficiently large. Also, we find that $N_{\mathrm{ESN}}$ does not limit the performance in linear tasks as the final readout layer can use the linear memory of the real-time QRC. In nonlinear tasks such as the determinant task increasing $N$ cannot compensate for a small $N_{\mathrm{ESN}}$; both layers become critical for a high performance.

\subsection{Performance advantage of deep hybrid design}

\begin{figure*}
    \includegraphics[trim=0cm 0cm 0cm 0cm,clip=true,width=0.95\textwidth]{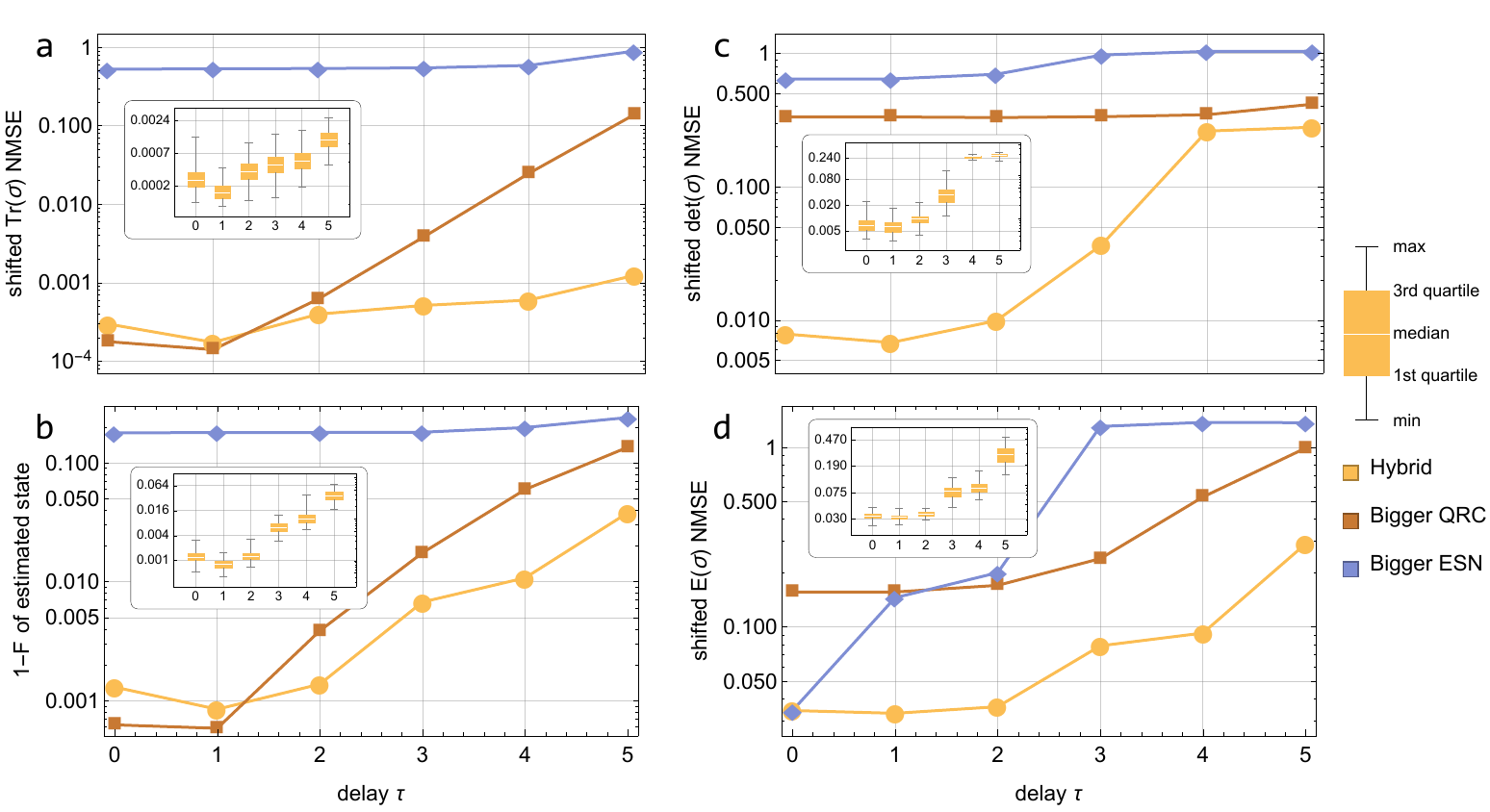}
    \caption{Mean performance of the hybrid architecture benchmarked against comparable purely quantum and purely classical alternatives. Shown are two linear tasks, the trace task of panel $(a)$ and state estimation task of panel $(b)$, and two nonlinear tasks, the determinant task of panel $(c)$ and entanglement detection tasks of panel $(d)$. Further details are given in Sec.~\ref{sec:tasks}. Insets show the hybrid performance in more detail. Shown results are averages over $100$ random realizations. Unlike in others in the trace task $\tau'=0$. All other parameters are as in Tab.~\ref{tab:parameters}.}
    \label{fig:lineresults}
\end{figure*}

Instead of using the hybrid reservoir computer, one could discard the ESN and use the output of the real-time QRC only. Alternatively, one could measure directly the $x$-quadratures of the available copies of $\sigma_t$ at each time step to form the estimate and use it with the ESN alone. Building on the results above, dropping either of the reservoirs is expected to hinder the performance because the real-time QRC is linear {in $\sigma_t$} the measurements are not informationally complete requiring the quantum pre-processing.  {Furthermore, as we have seen hybrid reservoir computing is not limited by the same SNR scaling of the QRC and can achieve comparable performance with a significantly smaller experimental overhead (less measurements).} In the next, we evaluate the performance of the hybrid architecture in comparison to fully quantum and classical reservoirs, for fixed output layer sizes.

In Fig.~\ref{fig:lineresults} we show the hybrid performance together with those of the real-time QRC and the ESN, in all the tasks considered before. While the Hilbert space offers a larger size than a classical reservoir with the same components \cite{PhysRevApplied.8.024030,nokkala2021gaussian}, here we do not make the assumption of an equal number of units. For a fair comparison, it is the size of $\mathbf{W}^{\mathrm{out}}$ to be kept fixed in all three cases, even if this requires a larger number of classical neurons. For the QRC, this is achieved by using two independent random instances, and for the ESN by simply doubling the number of neurons.
The insets show the hybrid performance in detail using a smaller scale. {Remarkably, the NMSE of the hybrid setup can be even three orders of magnitude better than the classical and quantum reservoirs, in both linear and nonlinear tasks. In all cases the hybrid performance in retrieving past states or their information (energy, purity, or entanglement) tends to be better or is at least comparable. This is a robust result and suggests that the hybrid advantage can be expected in general.}

In most cases the homodyne measurements alone do not reveal sufficient information to solve the tasks, causing the ESN to fail. The sole exception is the entanglement detection task; indeed, the covariance matrix of the used states has only two linearly independent elements which are both revealed by measuring only the $x$-quadrature as shown by Eq.~$(B2)$. Still, the ESN struggles to simultaneously handle the delay and the nonlinearity; presumably, this could be solved by increasing its size further. That being said, for generic bipartite states determining the logarithmic negativity requires the determinant of the blocks and the full covariance matrix and can therefore be expected to be impossible from measuring just the $x$-quadratures.

As expected, the real-time QRC can't cope with the nonlinear tasks. Furthemore, even in linear tasks, it is not competitive at higher delays due to the limited accuracy. The most striking difference is in the trace task where in fact the full delay is moved to the final readout layer by setting $\tau'=0$. This results in an error at delay $\tau=5$ that is approximately two orders of magnitude smaller for the hybrid model. As a note, in the state estimation task setting $\tau'=0$ would lead to worse performance for the hybrid and therefore the shown results use the default value ($\tau'=\lceil\tau/2\rceil$). This could be interpreted as an indication that the task is more difficult and hence it is better to share the delay.

\section{Conclusion and dicussion}

We have introduced a deep architecture for temporal quantum data processing by concatenating a scalable real-time quantum reservoir computer with a classical reservoir computer. The quantum layer is based on experimentally convenient Gaussian states and homodyne measurements, evades the measurement back-action, and is amenable to continuous extraction of data from a physical ensemble of identical random reservoirs. Our proposal combines its potential for a high information processing capacity and quantum preprocessing of the data with the nonlinear and versatile memory of the classical layer, leading to higher performance than either layer can achieve alone. Importantly, this performance advantage cannot be compensated for by just making the shallow counterparts bigger as it arises from the combination of their respective unique properties whereas training of the hybrid setup can be done by applying standard methods and is therefore of comparable ease. While the relatively small classical layers we have used are enough to support our main conclusions, they could be easily scaled up significantly; this would provide a further boost to the short-term memory without increasing the physical overhead, i.e. the ensemble size of the quantum layer. 

There is much room for further work. QRC has been proposed for temporal quantum data processing only recently \cite{mujal2021opportunities,tran2021learning,nokkala2023online,de2023quantum}, and indeed also the tasks considered here are benchmark tasks meant to be used as diagnostic tools to assess the capabilities of the reservoir computer rather than solve a practical problem. An interesting possibility to further explore comes from the memory capabilities of this design. Indeed, from single-time measurements, it is possible to gather, for instance, the entanglement evolution at several previous instants, offering a convenient alternative to continuous monitoring, where a final measurement can be used to reconstruct the dynamics. Identifying further use cases should however happen alongside proof-of-principle experiments. Our proposal is well suited for this purpose as it is within reach of state-of-the-art multimode quantum optics platforms. This would expand earlier experimental work on classical machine learning applied to tomographic tasks \cite{gebhart2023learning}, demonstrating for example improvements in speed and the number of measurements \cite{lennon2019efficiently} or the ability also to learn quantum trajectories \cite{flurin2020using}. Simulations have suggested, e.g., the possibility of significantly increased robustness to noisy or incomplete data \cite{lohani2020machine}. Another avenue of further research would be to consider the possibility of having  
temporal correlations or entanglement, i.e. between inputs at different time steps. This could increase the importance of the quantum memory to redistribute and store the relevant information such that it leaves its fingerprint on the extracted data processed by the classical layer. Generalization beyond bipartite inputs is another possibility. One may also use the deep architecture for classical temporal tasks as in previous proposals  where Gaussian states allow polynomial advantage \cite{nokkala2021gaussian, garcia2023scalable} in information processing capacity \cite{dambre2012information} over classical alternatives. Indeed, shallow (quantum) reservoirs have previously achieved high performance in, e.g., nonlinear channel equalization \cite{nokkala2021high}, nonlinear autoregressive moving average \cite{nokkala2021high,garcia2023squeezing} and chaotic time series prediction \cite{nokkala2021high,garcia2023scalable,garcia2023squeezing}. Finally, the possibility of augmenting the quantum layer with suitable non-Gaussian operations can be envisioned, combining the hybrid setup with recent advances in multimode single-photon-added and
-subtracted states of light \cite{ra2020non,biagi2020entangling} or considering single-photon instead of homodyne detection \cite{Spagnolo2022,dudas2023quantum} to unlock even more power.

\textit{Note added}: After the completion of this work, we
became aware of Ref.~\cite{wudarski2023hybrid} which also proposes combining a quantum and a classical reservoir computer, but to process classical temporal data.

\section*{Data availability statement} 

The code to reproduce the figures is available at https://github.com/jsinok/DeephybridQRC

\section*{Acknowledgements}

We acknowledge the Spanish State Research Agency, through the Mar\'ia de Maeztu project CEX2021-001164-M funded by the MCIN/AEI/10.13039/501100011033 and through the COQUSY project PID2022-140506NB-C21 and -C22 funded by MCIN/AEI/10.13039/501100011033, MINECO through the QUANTUM SPAIN project, and EU through the RTRP - NextGenerationEU within the framework of the Digital Spain 2025 Agenda. JN gratefully acknowledges financial support from the Academy of Finland under Project No. 348854.  GLG is funded by the Spanish  Ministerio de Educaci\'on y Formaci\'on Profesional/Ministerio de Universidades and co-funded by the University of the Balearic Islands through the Beatriz Galindo program (BG20/00085). 

\appendix

\section{\label{app:hypers}Hyperparameter optimization}

\begin{figure}[htbp]
\centering
    \includegraphics[width=0.45\textwidth]{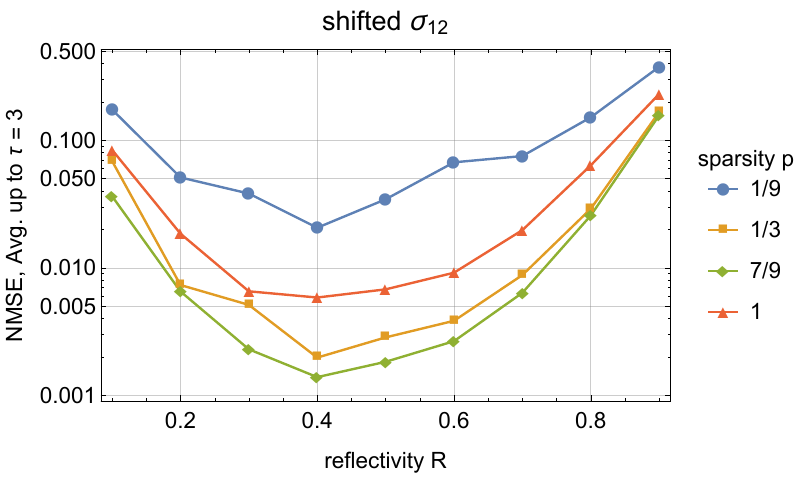}
    \caption{
    Effect of the value of reflectivity $R$ and the sparsity parameter $p$ on the real-time QRC performance. The task is to recall the off-diagonal elements of past input covariance matrices, shifted to have zero mean. $100$ random realizations have been considered for each value of $R$, $p$ and $\tau$.
    Other parameters are as in Tab.~I.}
    \label{fig:hyperparameters}
\end{figure}

\begin{figure}[htbp]
\centering
    \includegraphics[trim=0cm 0cm 0cm 0cm,clip=true,width=0.45\textwidth]{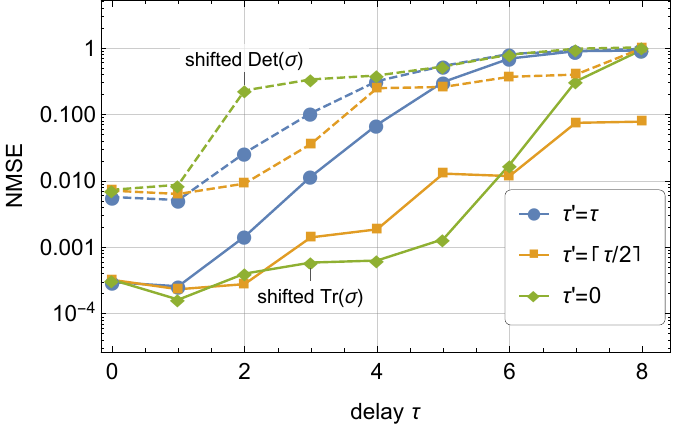}
    \caption{    
    Effect of different values of the delay $\tau'$ for the real-time QRC on mean performance over $100$ random realizations in both linear (shifted trace) and nonlinear (shifted determinant) tasks. At $\tau'=0$ ($\tau'=\tau$) all delay is handled by the ESN (real-time QRC). Other parameters are as in Tab.~I.     
    }
    \label{fig:delaysplit}
\end{figure}

By hyperparameters we mean any parameters of the hybrid setup other than the trained weights of matrices $\mathbf{W}^{\mathrm{out}}$ and $\mathbf{W}^{(2)}$. The number of optical modes $N$, neurons $N_{\mathrm{ESN}}$ and the ensemble size $M$ are special as performance can be expected to never decrease as they are increased due to some very general properties of reservoir computers \cite{dambre2012information} for the first two and for $M$ due to its connection to the SNR of the estimated covariance matrix $\tilde{\sigma}_x$.

Besides $N$ the real-time QRC dynamics depends on the reflectivity of the beam splitter $R$ and the sparsity parameter $p$.  We will analyze how they affect performance in a linear task of estimating the off-diagonal element of the input covariance matrix, averaged over small delays up to $\tau=3$ in Fig.~\ref{fig:hyperparameters}. Besides short-term memory, this also tests the ability of the quantum layer to reveal new information about the input state as normally measuring the $x$-quadrature can reveal only one of the diagonal elements. The performance behaves non-monotonically both in $R$ and $p$ and in particular, the tested short-term memory is optimized at intermediate values. The smallest error with the considered values is found at $R=0.4$ and $p=7/9$. Although considering longer delays would shift the optimal reflectivity to higher values we have limited to small delays as the rest can be taken care of by the ESN.

Adding the ESN introduces new parameters: besides $N_{\mathrm{ESN}}$ they are the separate delay for the real-time QRC $\tau'$, feedback gain $\rho$, and input gain $\iota$. The latter two are optimized with a simple lattice search separately for linear and nonlinear tasks. Unlike the others, $\tau'$ is a parameter unique to the hybrid setup. Its choice is investigated in Fig.~\ref{fig:delaysplit} considering the trace and determinant tasks. Differences become noticeable starting from $\tau=2$. In the linear trace task letting the classical layer handle the full delay leads to the smallest error until eventually the ESN runs out of memory and the NMSE collapses near $1$, leaving the even split to work best although still at a relatively high NMSE of $0.1$. But in the nonlinear determinant task, the roles are almost reversed, except none of the approaches produce useful results at larger delays. All in all, it seems that the default value $\tau'=\lceil\tau/2\rceil$ (with $ \lceil\cdot\rceil$ for the integer part)
tends to work well whereas $\tau'=\tau$ tends to lead to inferior performance, possibly because then the SNR of the memory provided by the real-time QRC is the lowest. For linear tasks trying also $\tau'=0$ may be worthwhile.

\section{\label{app:Gaussian}Gaussian quantum states and operations}

Gaussian states and operations are covered in detail for example in Refs.~\cite{ferraro2005gaussian,olivares2012quantum,serafini2017quantum}. Zero mean Gaussian states are completely determined by the covariance matrix which accounts for the shape of the multinormal distribution in quantum optical phase space, or the Wigner function of a Gaussian state. The covariance matrix of the vacuum state $\Ket{0}$ is normalized to the identity matrix. For a single mode squeezed thermal state with $n_{\mathrm{th}}$ thermal excitations, a squeezing parameter $r$ and phase of squeezing $\varphi$, the covariance matrix reads
\begin{equation}
    \sigma(n_{\mathrm{th}},r,\varphi)=(1+2n_{\mathrm{th}})\begin{pmatrix}
    y_{\cosh}+z_{\cos} & z_{\sin}\\
    z_{\sin} & y_{\cosh}-z_{\cos}
    \end{pmatrix}
    \label{eq:singlemodeinputmatrix}
\end{equation}
where $y_{\cosh}=\cosh{(2r)}$, $z_{\cos}=\cos{(\varphi)}\sinh{(2r)}$ and $z_{\sin}=\sin{(\varphi)}\sinh{(2r)}$. The energy $\langle H\rangle=\langle a^\dagger a\rangle+1/2$ and purity $\mu=(1+2n_{\mathrm{th}})^{-1}$ can be related to the trace and determinant, respectively, of the covariance matrix of Eq.~\eqref{eq:singlemodeinputmatrix}. Indeed, by direct calculation it can be seen that  $\mathrm{Tr}(\sigma(n_{\mathrm{th}},r,\varphi))=4(\langle a^\dagger a\rangle+1/2-\langle a\rangle \langle a^\dagger\rangle)=4\langle H\rangle$, where the last equality holds for zero mean states such as the ones considered here, and $\mathrm{Det}(\sigma(n_{\mathrm{th}},r,\varphi))=(1+2n_{\mathrm{th}})^2=1/\mu^2$.

Two-mode squeezed thermal states are considered in the entanglement detection task. The matrix for such a state with two-mode squeezing parameter $s$ reads
\begin{equation}
    \sigma(n_{\mathrm{th}},s)=(1+2n_{\mathrm{th}})\begin{pmatrix}
    y_{\cosh} & 0 & y_{\sinh} & 0 \\
    0 & y_{\cosh} & 0 & -y_{\sinh} \\
    y_{\sinh} & 0 & y_{\cosh} & 0 \\
    0 & -y_{\sinh} & 0 & y_{\cosh} \\
    \end{pmatrix}
    \label{eq:2modeinputmatrix}
\end{equation}
where now $y_{\cosh}=\cosh{(2s)}$ and $y_{\sinh}=\sinh{(2s)}$. The covariance matrix of a product state, such as the ones injected into the reservoir at each time step, is simply the direct sum of relevant covariance matrices. The initial state of the reservoir is also a product state, namely that of single-mode vacuum states for each reservoir mode. The corresponding covariance matrix is just an identity matrix of appropriate size.

Unitary evolution is determined by a Hamiltonian $H$. Let $\mathbf{M}$ be the real symmetric matrix defined by
\begin{equation}
    H=\frac{1}{2}\mathbf{x}^\top\mathbf{M}\mathbf{x}
\end{equation}
where $\mathbf{x}=\{x_1,p_1,x_2,p_2,\ldots\}^\top$ is the vector of the quadrature operators. Then the symplectic matrix $\mathbf{S}(t)$ that determines the evolution of the covariance matrix $\sigma$ via $\sigma(t)=\mathbf{S}(t)\sigma(\mathbf{S}(t))^\top$ is given by
\begin{equation}
\mathbf{S}(t)=\exp(\mathbf{\Omega}\mathbf{M}t)
\end{equation}
where $\mathbf{\Omega}_{jl}=-\mathrm{i}[\mathbf{x}_j,\mathbf{x}_l]$ is the symplectic form accounting for the commutation relations between the quadrature operators. 

In the case at hand it is convenient to consider separately the two nonlinear crystals with Hamiltonians $H_1$ and $H_2$ and the beam splitter with reflectivity $R$. The symplectic matrix $\mathbf{S}(\Delta t)$ that accounts for the evolution during a single timestep reads
\begin{equation}
    \mathbf{S}(\Delta t)=\begin{pmatrix}
        \sqrt{R}\mathbf{S}_1(\Delta t) & -\sqrt{1-R}\mathbf{S}_1(\Delta t) \\
        \sqrt{1-R}\mathbf{S}_2(\Delta t) & \sqrt{R}\mathbf{S}_2(\Delta t)
    \end{pmatrix}
    \label{eq:symplecticmatrix}
\end{equation}
where $\mathbf{S}_{i}(\Delta t)=\exp(\mathbf{\Omega}\mathbf{M}_i\Delta t)$ and $\Delta t$ is the time spent inside the first or the second crystal. Importantly, the first (second) column of this block matrix acts on the reservoir (ancillary) modes and in general, redistributes the information such that both the reservoir state and the output will depend on input history as is necessary to achieve reservoir computing. After each action of $\mathbf{S}(\Delta t)$ the $x$-quadratures of the ancillary modes will be measured and then replaced by fresh modes accounting for the new input; in practice, the correlation blocks between the reservoir and ancillary modes are set to the zero matrix and the ancillary block to the new covariance matrix.

In general, the beam splitter will create correlations between the reservoir and ancillary modes which will lead to measurement back-action conditioned by the measurement outcomes.  However, as shown in the appendices of Ref.~\cite{garcia2023scalable},  here the back-action averages out and one gets equivalent results by propagating the back-action free reservoir covariance matrix determined only by the Hamiltonian evolution and input history.

It remains to model the measurement process given a finite sample size $M$ and to construct the maximum likelihood estimate $\Tilde{\sigma}_{\mathbf{x}}$ of the covariance matrix $\sigma_{\mathbf{x}}$ of the $x$-quadratures of the ancillary modes. It is convenient to combine these two steps into one by considering the sampling distribution of such a process, i.e. the Wishart distribution $W(\sigma_{\mathbf{x}},M)$ \cite{wishart1928generalised}. Since
\begin{equation}
    \Tilde{\sigma}_{\mathbf{x}}\sim\frac{1}{M}W(\sigma_{\mathbf{x}},M),
    \label{eq:wishart}
\end{equation}
it suffices to draw a random sample from $W(\sigma_{\mathbf{x}},M)$ to account for stochasticity caused by finite ensemble effects. For this, it is enough to sample suitable chi-squared distributions and the normal distribution of zero mean and unit variance, as well as find the Cholesky decomposition of $\sigma_{\mathbf{x}}$ \cite{smith1972algorithm}. In the limit $M\rightarrow\infty$ one naturally recovers $\Tilde{\sigma}_{\mathbf{x}}\rightarrow\sigma_{\mathbf{x}}$, and in this limit the sampling can be skipped.

For the state estimation task, the target is a covariance matrix and both the fidelity $F$ and the infidelity $1-F$ between the actual and estimated state are used as the figure of merit. In terms of some density operators $\varrho$ and $\varsigma$ the definition reads
\begin{equation}
F(\varrho,\varsigma)=\left(\mathrm{Tr}\sqrt{\sqrt{\varrho}\varsigma\sqrt{\varrho}}\right)^2
\label{eq:fidelity}
\end{equation}
and the value can be found directly from the covariance matrices \cite{olivares2012quantum}. It should be noted that since the training minimizes the squared errors between their elements it does not necessarily optimize fidelity. However, low errors tend to result in high fidelities, as might be expected.

Logarithmic negativity $E(\sigma)$ is estimated in the entanglement detection task. It is based on the positive partial transpose criterion and quantifies how much the positivity of the partial transpose is violated, or how much the purity of the state would need to be decreased to make it separable. In the case at hand it is determined by $\tilde{d}_{-}$, the smaller of the two symplectic eigenvalues of the two-mode covariance matrix $\sigma$ \cite{PhysRevA.72.032334}, via
\begin{equation}
    \mathrm{E}(\sigma)=\max\{0,-\log_2{\tilde{d}_{-}}\}.
\end{equation}
In the training phase, the target is $-\log_2{\tilde{d}_{-}}$, and the maximization is done separately before calculating the NMSE. It should be pointed out that in many sources the relevant quantity is reported to be $-\log_2{2\tilde{d}_{-}}$ but this constant factor difference is due to a different normalization of the vacuum covariance matrix; the definition introduced above is equivalent.

\section{\label{app:RC}Reservoir computing}

Generally speaking, a reservoir computer is a dynamical input-output system such that under the drive of the input, its dynamical variables become diverse functions of input history \cite{konkoli2017reservoir,grigoryeva2018echo,nakajima2021reservoir}. The output is trained to approximate the reservoir variables; typically linear combinations  are used to offload the bulk of the computations to the system, i.e. the reservoir. This leads to a supervised machine learning approach well suited for temporal tasks, as recurrent neural networks. More formally, a reservoir computer stands on the echo state  \cite{jaeger2001echo} and fading memory \cite{boyd1985fading} properties, which essentially guarantee that eventually the effect of the initial conditions becomes negligible and the dynamical variables become continuous functions of only recent input history--- so-called fading memory functions---respectively \cite{konkoli2017reservoir}. The real-time QRC in particular is known to have these properties \cite{garcia2023scalable}. Reservoir computers are ideal for solving tasks that can be well approximated by fading memory functions and have the advantage of robust single-step training and amenability to physical implementations \cite{tanaka2019recent} when compared to recurrent neural networks. The dynamical variables used to form the output are the computational nodes. In our case, they are the elements of the estimated covariance matrix of the $x$-quadratures $\tilde{\sigma}_x$ and the neurons for the real-time QRC and the ESN, respectively.

During the training and testing phases, the states of the computational nodes are recorded into two matrices $\mathbf{X}_{\mathrm{train}}$ and $\mathbf{X}_{\mathrm{test}}$ such that the rows correspond to time steps and the columns to each computational node. The bias term is included by adding a unit column. Let $\Bar{\mathbf{o}}_{\mathrm{train}}$ be the target output in the training phase. Then we set for example
\begin{equation}   
\mathbf{W}^{\mathrm{out}}=\mathbf{X}_{\mathrm{train}}^+\Bar{\mathbf{o}}^{\top}_{\mathrm{train}}
\end{equation}
where $\mathbf{X}_{\mathrm{train}}^+$ is the Moore–Penrose pseudoinverse of $\mathbf{X}_{\mathrm{train}}$, minimizing the squared error $\sum_i(\mathbf{o}_{\mathrm{train},i}-\Bar{\mathbf{o}}_{\mathrm{train},i})^2$ between the actual output $\mathbf{o}_{\mathrm{train}}=\mathbf{W}^{\mathrm{out}\top}\mathbf{X}_{\mathrm{train}}^\top$ and the target $\Bar{\mathbf{o}}_{\mathrm{train}}$. Test phase uses the trained weights to generate the output $\mathbf{o}_{\mathrm{test}}=\mathbf{X}_{\mathrm{test}}\mathbf{W}^{\mathrm{out}}$. The standard figure of merit is the normalized mean squared error (NMSE), namely
\begin{equation}
    \mathrm{NMSE}(\mathbf{o}_{\mathrm{test}},\Bar{\mathbf{o}}_{\mathrm{test}})=\frac{\sum_i(\mathbf{o}_{\mathrm{test},i}-\Bar{\mathbf{o}}_{\mathrm{test},i})^2}{\sum_i\Bar{\mathbf{o}}_{\mathrm{test},i}^2}.
\label{eq:NMSE}
\end{equation}
 {In the short-term memory task of quantum states, however, fidelity is used.} Importantly, NMSE in the test phase depends on the ability of the reservoir computer to generalize beyond the training phase.
Further remarks are in order. To account for the hybrid architecture shown in Fig.~1 of the main text, the real time QRC is trained first. Its trained output is then treated as input for the ESN and training is repeated for the full hybrid setup. When the target is a vector, the framework above applies as is and in particular the trained matrix will have its own column for each element while the squared error is minimized element-wise. In all tasks where the final figure of merit is NMSE, the final target is also a scalar, making Eq.~\eqref{eq:NMSE} sufficient.

\section{\label{app:randomgeneration}Generation of random instances}

The relevant parameters take the values given in Tab.~I unless stated otherwise.
Each random instance involves the generation of an input time series where elements are sampled according to either Eq.~$(4)$ or $(5)$, the generation of the random Hamiltonians $H_{\chi^{(2)}}$ for the two nonlinear crystals, the weight matrix $\mathbf{W}$ for the ESN and the coercion matrix $\mathbf{C}$. It should be noted that the Hamiltonians $H_{\chi^{(2)}}$ are not by construction stable; in the rare event that they are not, they are replaced by new random Hamiltonians. The relevance of stability for the real-time QRC is discussed in App.~F of \cite{garcia2023scalable}. The target time series is completely determined by the input in all tasks. As previously explained, the only effect of the finite ensemble size $M$ is to introduce some random noise in the observables of the real-time QRC; this is accounted for by using the sampling distribution (see Eq.~\eqref{eq:wishart}) once for each time step.

\section*{References}

\normalem
\bibliographystyle{iopart-num}
\bibliography{references}

\providecommand{\newblock}{}
\begin{thebibliography}{10}
\expandafter\ifx\csname url\endcsname\relax
  \def\url#1{{\tt #1}}\fi
\expandafter\ifx\csname urlprefix\endcsname\relax\def\urlprefix{URL }\fi
\providecommand{\eprint}[2][]{\url{#2}}

\bibitem{hinton2006fast}
Hinton G~E, Osindero S and Teh Y~W 2006 {\em Neural computation\/} {\bf 18} 1527--1554

\bibitem{Goodfellow2016}
Goodfellow I, Bengio Y and Courville A 2016 {\em Deep Learning\/} (MIT Press) \url{http://www.deeplearningbook.org}

\bibitem{farhi2014quantum}
Farhi E, Goldstone J and Gutmann S 2014 {\em arXiv preprint arXiv:1411.4028\/}

\bibitem{kandala2017hardware}
Kandala A, Mezzacapo A, Temme K, Takita M, Brink M, Chow J~M and Gambetta J~M 2017 {\em nature\/} {\bf 549} 242--246

\bibitem{mitarai2018quantum}
Mitarai K, Negoro M, Kitagawa M and Fujii K 2018 {\em Physical Review A\/} {\bf 98} 032309

\bibitem{zhu2019training}
Zhu D, Linke N~M, Benedetti M, Landsman K~A, Nguyen N~H, Alderete C~H, Perdomo-Ortiz A, Korda N, Garfoot A, Brecque C {\em et~al.\/} 2019 {\em Science advances\/} {\bf 5} eaaw9918

\bibitem{verdon2019learning}
Verdon G, Broughton M, McClean J~R, Sung K~J, Babbush R, Jiang Z, Neven H and Mohseni M 2019 {\em arXiv preprint arXiv:1907.05415\/}

\bibitem{mujal2021opportunities}
Mujal P, Mart{\'\i}nez-Pe{\~n}a R, Nokkala J, Garc{\'\i}a-Beni J, Giorgi G~L, Soriano M~C and Zambrini R 2021 {\em Advanced Quantum Technologies\/} {\bf 4} 2100027

\bibitem{cerezo2021variational}
Cerezo M, Arrasmith A, Babbush R, Benjamin S~C, Endo S, Fujii K, McClean J~R, Mitarai K, Yuan X, Cincio L {\em et~al.\/} 2021 {\em Nature Reviews Physics\/} {\bf 3} 625--644

\bibitem{bharti2022noisy}
Bharti K, Cervera-Lierta A, Kyaw T~H, Haug T, Alperin-Lea S, Anand A, Degroote M, Heimonen H, Kottmann J~S, Menke T {\em et~al.\/} 2022 {\em Reviews of Modern Physics\/} {\bf 94} 015004

\bibitem{ghukasyan2023quantum}
Ghukasyan A, Baker J~S, Goktas O, Carrasquilla J and Radha S~K 2023 {\em arXiv preprint arXiv:2305.17707\/}

\bibitem{peruzzo2014variational}
Peruzzo A, McClean J, Shadbolt P, Yung M~H, Zhou X~Q, Love P~J, Aspuru-Guzik A and O’brien J~L 2014 {\em Nature communications\/} {\bf 5} 4213

\bibitem{Chen_2021}
Chen S~Y~C, Huang C~M, Hsing C~W and Kao Y~J 2021 {\em Machine Learning: Science and Technology\/} {\bf 2} 045021 \urlprefix\url{https://dx.doi.org/10.1088/2632-2153/ac104d}

\bibitem{PhysRevApplied.8.024030}
Fujii K and Nakajima K 2017 {\em Phys. Rev. Appl.\/} {\bf 8}(2) 024030 \urlprefix\url{https://link.aps.org/doi/10.1103/PhysRevApplied.8.024030}

\bibitem{martinez2021dynamical}
Mart{\'\i}nez-Pe{\~n}a R, Giorgi G~L, Nokkala J, Soriano M~C and Zambrini R 2021 {\em Physical Review Letters\/} {\bf 127} 100502

\bibitem{tran2021learning}
Tran Q~H and Nakajima K 2021 {\em Physical review letters\/} {\bf 127} 260401

\bibitem{ghosh2019quantum}
Ghosh S, Opala A, Matuszewski M, Paterek T and Liew T~C 2019 {\em npj Quantum Information\/} {\bf 5} 35

\bibitem{PhysRevLett.123.260404}
Ghosh S, Paterek T and Liew T~C~H 2019 {\em Phys. Rev. Lett.\/} {\bf 123}(26) 260404 \urlprefix\url{https://link.aps.org/doi/10.1103/PhysRevLett.123.260404}

\bibitem{suprano2023experimental}
Suprano A, Zia D, Innocenti L, Lorenzo S, Cimini V, Giordani T, Palmisano I, Polino E, Spagnolo N, Sciarrino F, Palma G~M, Ferraro A and Paternostro M 2024 {\em Phys. Rev. Lett.\/} {\bf 132}(16) 160802 \urlprefix\url{https://link.aps.org/doi/10.1103/PhysRevLett.132.160802}

\bibitem{Krisnanda}
Krisnanda T, Paterek T, Paternostro M and Liew T~C 2023 {\em Physical Review D\/} {\bf 107} 086014

\bibitem{Innocenti2023}
Innocenti L, Lorenzo S, Palmisano I, Ferraro A, Paternostro M and Palma G~M 2023 {\em Communications Physics\/} {\bf 6} 118 ISSN 2399-3650 \urlprefix\url{https://doi.org/10.1038/s42005-023-01233-w}

\bibitem{jaeger2001echo}
Jaeger H 2001 {\em Bonn, Germany: German National Research Center for Information Technology GMD Technical Report\/} {\bf 148} 13

\bibitem{maass2004computational}
Maass W and Markram H 2004 {\em Journal of computer and system sciences\/} {\bf 69} 593--616

\bibitem{nakajima2021reservoir}
Nakajima K and Fischer I 2021 {\em Reservoir Computing\/} (Springer)

\bibitem{voulodimos2018deep}
Voulodimos A, Doulamis N, Doulamis A and Protopapadakis E 2018 {\em Computational intelligence and neuroscience\/} {\bf 2018}

\bibitem{young2018recent}
Young T, Hazarika D, Poria S and Cambria E 2018 {\em IEEE Computational Intelligence Magazine\/} {\bf 13} 55--75

\bibitem{koutny2023deep}
Koutn{\`y} D, Gin{\'e}s L, Mocza{\l}a-Dusanowska M, H{\"o}fling S, Schneider C, Predojevi{\'c} A and Je{\v{z}}ek M 2023 {\em Science Advances\/} {\bf 9} eadd7131

\bibitem{triefenbach2010phoneme}
Triefenbach F, Jalalvand A, Schrauwen B and Martens J~P 2010 {\em Advances in neural information processing systems\/} {\bf 23}

\bibitem{gallicchio2017deep}
Gallicchio C, Micheli A and Pedrelli L 2017 {\em Neurocomputing\/} {\bf 268} 87--99

\bibitem{freiberger2019improving}
Freiberger M, Sackesyn S, Ma C, Katumba A, Bienstman P and Dambre J 2019 {\em IEEE Journal of Selected Topics in Quantum Electronics\/} {\bf 26} 1--11

\bibitem{nakajima2022physical}
Nakajima M, Inoue K, Tanaka K, Kuniyoshi Y, Hashimoto T and Nakajima K 2022 {\em Nature Communications\/} {\bf 13} 7847

\bibitem{lin2022deep}
Lin B~D, Shen Y~W, Tang J~Y, Yu J, He X and Wang C 2022 {\em IEEE Journal of Selected Topics in Quantum Electronics\/} {\bf 29} 1--8

\bibitem{lupo2023deep}
Lupo A, Picco E, Zajnulina M and Massar S 2023 {\em Optica\/} {\bf 10} 1478--1485

\bibitem{nokkala2023online}
Nokkala J 2023 {\em Scientific Reports\/} {\bf 13} 7694

\bibitem{tanaka2019recent}
Tanaka G, Yamane T, H{\'e}roux J~B, Nakane R, Kanazawa N, Takeda S, Numata H, Nakano D and Hirose A 2019 {\em Neural Networks\/} {\bf 115} 100--123

\bibitem{chen2020temporal}
Chen J, Nurdin H~I and Yamamoto N 2020 {\em Physical Review Applied\/} {\bf 14} 024065

\bibitem{sannia2022dissipation}
Sannia A, Mart{\'\i}nez-Pe{\~n}a R, Soriano M~C, Giorgi G~L and Zambrini R 2024 {\em Quantum\/} {\bf 8} 1291

\bibitem{PhysRevResearch.5.023057}
Kubota T, Suzuki Y, Kobayashi S, Tran Q~H, Yamamoto N and Nakajima K 2023 {\em Phys. Rev. Res.\/} {\bf 5}(2) 023057 \urlprefix\url{https://link.aps.org/doi/10.1103/PhysRevResearch.5.023057}

\bibitem{domingo2023taking}
Domingo L, Carlo G and Borondo F 2023 {\em Scientific Reports\/} {\bf 13} 8790

\bibitem{mujal2023time}
Mujal P, Mart{\'\i}nez-Pe{\~n}a R, Giorgi G~L, Soriano M~C and Zambrini R 2023 {\em npj Quantum Information\/} {\bf 9} 16

\bibitem{garcia2023scalable}
Garc{\'\i}a-Beni J, Giorgi G~L, Soriano M~C and Zambrini R 2023 {\em Physical Review Applied\/} {\bf 20} 014051

\bibitem{garcia2023squeezing}
Garc\'{i}a-Beni J, Giorgi G~L, Soriano M~C and Zambrini R 2024 {\em Opt. Express\/} {\bf 32} 6733--6747 \urlprefix\url{https://opg.optica.org/oe/abstract.cfm?URI=oe-32-4-6733}

\bibitem{yasuda2023quantum}
Yasuda T, Suzuki Y, Kubota T, Nakajima K, Gao Q, Zhang W, Shimono S, Nurdin H~I and Yamamoto N 2023 {\em arXiv preprint arXiv:2310.06706\/}

\bibitem{innocenti}
Innocenti L, Lorenzo S, Palmisano I, Ferraro A, Paternostro M and Palma G~M 2023 {\em Communications Physics\/} {\bf 6} 118

\bibitem{tureciPRX23}
Hu F, Angelatos G, Khan S~A, Vives M, T{\"u}reci E, Bello L, Rowlands G~E, Ribeill G~J and T{\"u}reci H~E 2023 {\em Physical Review X\/} {\bf 13} 041020

\bibitem{konkoli2017reservoir}
Konkoli Z 2017 On reservoir computing: from mathematical foundations to unconventional applications {\em Advances in unconventional computing\/} (Springer) pp 573--607

\bibitem{de2023quantum}
De~Prins R, Van~der Sande G and Bienstman P 2023 {\em arXiv preprint arXiv:2306.00134\/}

\bibitem{nokkalareview}
Nokkala J, Piilo J and Bianconi G 2023 {\em arXiv preprint arXiv:2311.16265\/}

\bibitem{renaultPRXquantum}
Renault P, Nokkala J, Roeland G, Joly N, Zambrini R, Maniscalco S, Piilo J, Treps N and Parigi V 2023 {\em PRX Quantum\/} {\bf 4}(4) 040310 \urlprefix\url{https://link.aps.org/doi/10.1103/PRXQuantum.4.040310}

\bibitem{pfeffer2022hybrid}
Pfeffer P, Heyder F and Schumacher J 2022 {\em Physical Review Research\/} {\bf 4} 033176

\bibitem{vidal2002computable}
Vidal G and Werner R~F 2002 {\em Physical Review A\/} {\bf 65} 032314

\bibitem{nokkala2021gaussian}
Nokkala J, Mart{\'\i}nez-Pe{\~n}a R, Giorgi G~L, Parigi V, Soriano M~C and Zambrini R 2021 {\em Communications Physics\/} {\bf 4} 53

\bibitem{gebhart2023learning}
Gebhart V, Santagati R, Gentile A~A, Gauger E~M, Craig D, Ares N, Banchi L, Marquardt F, Pezz{\`e} L and Bonato C 2023 {\em Nature Reviews Physics\/} {\bf 5} 141--156

\bibitem{lennon2019efficiently}
Lennon D~T, Moon H, Camenzind L~C, Yu L, Zumb{\"u}hl D~M, Briggs G~A~D, Osborne M~A, Laird E~A and Ares N 2019 {\em npj Quantum Information\/} {\bf 5} 79

\bibitem{flurin2020using}
Flurin E, Martin L~S, Hacohen-Gourgy S and Siddiqi I 2020 {\em Physical Review X\/} {\bf 10} 011006

\bibitem{lohani2020machine}
Lohani S, Kirby B~T, Brodsky M, Danaci O and Glasser R~T 2020 {\em Machine Learning: Science and Technology\/} {\bf 1} 035007

\bibitem{dambre2012information}
Dambre J, Verstraeten D, Schrauwen B and Massar S 2012 {\em Scientific reports\/} {\bf 2} 514

\bibitem{nokkala2021high}
Nokkala J, Mart{\'\i}nez-Pe{\~n}a R, Zambrini R and Soriano M~C 2022 {\em IEEE Transactions on Neural Networks and Learning Systems\/} {\bf 33} 2664--2675

\bibitem{ra2020non}
Ra Y~S, Dufour A, Walschaers M, Jacquard C, Michel T, Fabre C and Treps N 2020 {\em Nature Physics\/} {\bf 16} 144--147

\bibitem{biagi2020entangling}
Biagi N, Costanzo L~S, Bellini M and Zavatta A 2020 {\em Physical Review Letters\/} {\bf 124} 033604

\bibitem{Spagnolo2022}
Spagnolo M, Morris J, Piacentini S, Antesberger M, Massa F, Crespi A, Ceccarelli F, Osellame R and Walther P 2022 {\em Nature Photonics\/} {\bf 16} 318--323 ISSN 1749-4893 \urlprefix\url{https://doi.org/10.1038/s41566-022-00973-5}

\bibitem{dudas2023quantum}
Dudas J, Carles B, Plouet E, Mizrahi F~A, Grollier J and Markovi{\'c} D 2023 {\em npj Quantum Information\/} {\bf 9} 64

\bibitem{wudarski2023hybrid}
Wudarski F, OConnor D, Geaney S, Asanjan A~A, Wilson M, Strbac E, Lott P~A and Venturelli D 2023 {\em arXiv preprint arXiv:2311.14105\/}

\bibitem{ferraro2005gaussian}
Ferraro A, Olivares S and Paris M~G 2005 {\em Gaussian States in Quantum Information\/} Napoli Series on physics and Astrophysics (Bibliopolis) ISBN 88-7088-483-X

\bibitem{olivares2012quantum}
Olivares S 2012 {\em The European Physical Journal Special Topics\/} {\bf 203} 3--24

\bibitem{serafini2017quantum}
Serafini A 2017 {\em Quantum continuous variables: a primer of theoretical methods\/} (CRC press)

\bibitem{wishart1928generalised}
Wishart J 1928 {\em Biometrika\/}  32--52

\bibitem{smith1972algorithm}
Smith W and Hocking R 1972 {\em Journal of the Royal Statistical Society. Series C (Applied Statistics)\/} {\bf 21} 341--345

\bibitem{PhysRevA.72.032334}
Adesso G and Illuminati F 2005 {\em Phys. Rev. A\/} {\bf 72}(3) 032334 \urlprefix\url{https://link.aps.org/doi/10.1103/PhysRevA.72.032334}

\bibitem{grigoryeva2018echo}
Grigoryeva L and Ortega J~P 2018 {\em Neural Networks\/} {\bf 108} 495--508

\bibitem{boyd1985fading}
Boyd S and Chua L 1985 {\em IEEE Transactions on circuits and systems\/} {\bf 32} 1150--1161

\end{thebibliography}
\end{document}